\documentclass[aps,prl,floatfix,twocolumn,superscriptaddress,tightenlines]{revtex4}

\usepackage{amsmath}
\usepackage{graphicx}
\usepackage{epstopdf}
\usepackage{psfrag}
\usepackage[usenames]{color}
\usepackage{bm}
\usepackage{soul}

\newcommand{\ket}[1]{\vert#1\rangle}
\newcommand{\bra}[1]{\langle#1\vert}

\DeclareFontFamily{U}{euc}{}
\DeclareFontShape{U}{euc}{m}{n}{<-6>eurm5<6-8>eurm7<8->eurm10}{}%
\DeclareSymbolFont{AMSc}{U}{euc}{m}{n}
\DeclareMathSymbol{\umu}{\mathord}{AMSc}{"16} 

\begin{document}

\title{Discrete single-photon quantum walks with tunable decoherence}

\author{M. A. Broome}
\affiliation{Department of Physics and Centre for Quantum Computer Technology, University of Queensland, Brisbane 4072, Australia}

\author{A. Fedrizzi}
\affiliation{Department of Physics and Centre for Quantum Computer Technology, University of Queensland, Brisbane 4072, Australia}

\author{B. P. Lanyon}
\affiliation{Department of Physics and Centre for Quantum Computer Technology, University of Queensland, Brisbane 4072, Australia}

\author{I. Kassal}
\affiliation{Department of Chemistry and Chemical Biology, Harvard University, Cambridge Massachusetts 02138, USA}

\author{A. Aspuru-Guzik}
\affiliation{Department of Chemistry and Chemical Biology, Harvard University, Cambridge Massachusetts 02138, USA}

\author{A. G. White}
\affiliation{Department of Physics and Centre for Quantum Computer Technology, University of Queensland, Brisbane 4072, Australia}

\date{\today}
\begin{abstract}
Quantum walks have a host of applications, ranging from quantum computing to the simulation of biological systems. We present an intrinsically stable, deterministic implementation of discrete quantum walks with single photons in space. The number of optical elements required scales linearly with the number of steps. We measure walks with up to $6$ steps and explore the quantum-to-classical transition by introducing tunable decoherence.  Finally, we also investigate the effect of absorbing boundaries and show that decoherence significantly affects the probability of absorption.
\end{abstract}

\maketitle

The random walk is a fundamental model of dynamical processes that has found extensive application in science.  The quantum walk (QW) is the extension of the random walk to the quantum regime~\cite{aharonov_quantum_1993, kempe_quantum_2003}. Here, the classical walker is replaced by a quantum particle, such as an electron, atom or photon, and the stochastic evolution by a unitary process. A key difference is that the many possible paths of the quantum walker can exhibit interference, leading to a very different probability distribution for finding the walker at a given location.

An important motivation for work on QW's has been their application to quantum computation: not only were they instrumental to Feynman's original quantum computer (as the clock mechanism~\cite{feynman_quantum_1986}), but it has since been shown that they represent a universal computational primitive~\cite{childs_universal_2008,lovett_universal_2009} and have inspired novel quantum algorithms~\cite{childs_exponential_2002,shenvi_quantum_2003,ambainis_quantum_2003}. QW's have also been used to analyze energy transport in biological systems~\cite{mohseni_environment-assisted_2008, rebentrost_environment-assisted_2009}.

Despite a few early experimental demonstrations ~\cite{bouwmeester_optical_1999, ryan2005eid, Do:05,ribeiro2008qrw}, experimentalists have only recently begun to develop the level of control over single quantum particles required to implement discrete-time QW's,  leading to demonstrations with neutral atoms in position space \cite{MichalKarski07102009}, ions in phase space~\cite{zaehringer2009rqw, PhysRevLett.103.090504}, and single photons in time~\cite{schreiber-2009}. Continuous-time quantum walks have different outcomes, applications, and experimental challenges, see, e.g., \cite{bromberg_quantum_2009}.

In this work, we present discrete-time QW's of single photons in space. Our approach is robust, due to the use of intrinsically stable interferometers, yet highly versatile---enabling control over every operation at every step of the walk. Of particular interest is the ability to introduce a controlled amount of decoherence, which we use to explore the quantum-to-classical transition. Besides being of fundamental interest, decoherence in QW's can improve the performance of certain computational protocols~\cite{kendon_decoherence_2003} and is particularly important in their application to describing energy transport~\cite{mohseni_environment-assisted_2008,rebentrost_environment-assisted_2009}. Finally, we investigate the effect of introducing absorbing boundaries into the walk, as theoretically investigated by~\cite{ambainis_one-dimensional_2001,bach_one-dimensional_2004}. 

The simplest random walk occurs on a one-dimensional lattice. The particle begins at one site and each step of the walk consists of a move to a neighboring site on the left or right, determined by the outcome of a coin flip. In the analogous QW, the coin is represented by a two-level quantum system whose orthogonal levels we will designate $\ket{H}$ (horizontal) and $\ket{V}$ (vertical). Each step of the QW starts with an analogue of the coin flip: a unitary \textit{coin operator} $C$ is applied to the coin space. An unbiased, or Hadamard, coin operator transforms the coin so that $\ket{H}{\rightarrow}\ket{D}{=}(\ket{H}{+}\ket{V})/\sqrt{2}$, $\ket{V}{\rightarrow}\ket{A}{=}(\ket{H}{-}\ket{V})/\sqrt{2}$. Following each coin operation is the \textit{shift operator}, 
\begin{equation}
S=\sum_j \ket{j-1}\bra{j}\otimes \ket{H}\bra{H} + \ket{j+1}\bra{j}\otimes \ket{V}\bra{V},
\label{eq:shift}
\end{equation}
which moves the particles to one of the neighboring lattice sites, conditional on the quantum coin state. Therefore, the operation $W=SC$ makes up a single step of the QW, and a walker in an initial state $\ket{\psi}$ is found in the state $W^N\ket{\psi}$ after $N$ steps.

The quantum and random walks can be considered the extremes of a spectrum, with pure quantum evolution turning into classical evolution if there is sufficient decoherence~
\cite{brun_quantum_2003,kendon_complementarity_2005,kendon_decoherence_2006}. In addition, all the intermediate walks are special cases of the broad category of quantum stochastic walks~\cite{rodriguezrosario-2009}. Various mechanisms of decoherence have been studied~\cite{kendon_decoherence_2006}; here we consider pure dephasing, since this corresponds to the decoherence mechanism in our experiment. The system is described by a density matrix $\rho$ which at each step undergoes the evolution
\begin{equation}
\label{eq:deco}
\rho_{N+1}=(1-q)W\rho_{N}W^{\dagger}+q\sum_i K_i W\rho_{N}W^{\dagger} K_{i}^{\dagger},
\end{equation}
where the Kraus operators $K_i=\ket{i}\bra{i}$ correspond to pure dephasing. The parameter $q$ is the probability of a dephasing event occuring at each step. If $q=0$, the walk is a pure QW, while $q=1$ reproduces the random walk. Note that to observe the quantum-to-classical transition, gradual decoherence must be applied at each step, not just to the initial coin state \cite{brun_quantum_2003,kendon_complementarity_2005, kendon_decoherence_2006}.

\begin{figure}[t]
\includegraphics[width=1 \columnwidth]{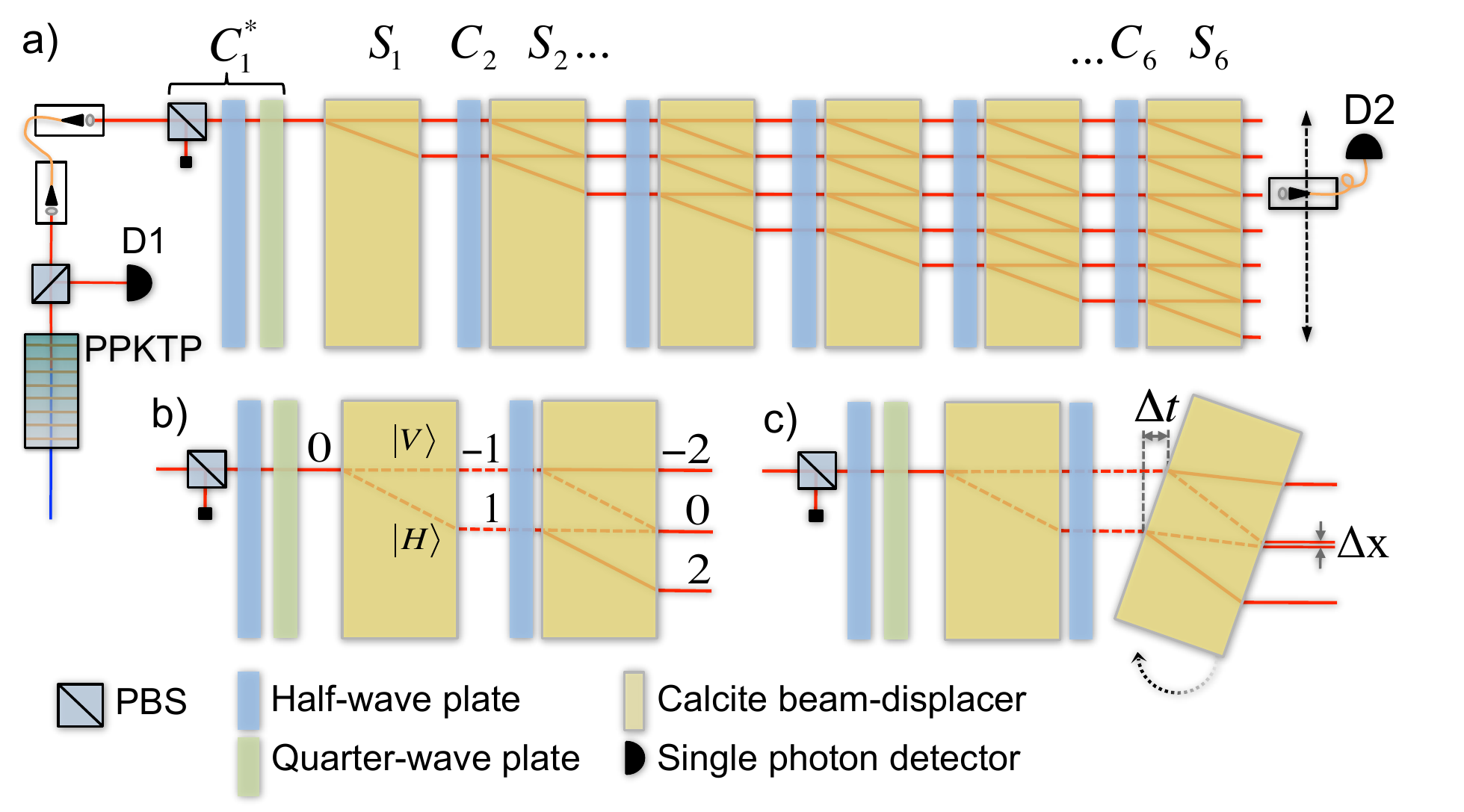}
\caption{\label{fig:scheme} Experimental schematic. 
(a) Photons created via spontaneous parametric down-conversion in a PPKTP crystal are injected into free-space mode 0. Arbitrary initial coin (polarization) states are prepared by the first polarizing beam splitter (PBS) and waveplate combination. Six pairs of coin ($C_{i}$) and shift ($S_{i}$) operators implement six steps of the walk. Coincident detection of photons at detector D2 and D1 (4.4 ns time window) herald a successful run of the walk. 
(b) Details of our optical mode numbering convention for the first two steps. The dashed lines trace out one of the interferometers which are used to align the quantum walk. 
(c) A relative angle between two beam displacers reduces the recombined photon's temporal ($\Delta t$) and spatial ($\Delta x$) mode overlap, thereby implementing tunable decoherence.}
\end{figure}

Our implementation of the discrete QW on a line represents a polarization analogue of the originally proposed linear-optical version of the Galton board based on beamsplitters and phaseshifters~\cite{jeong2004sqr, Do:05}. A similar, polarization-encoded setup has been proposed for cube polarizing beamsplitters \cite{zhao2002iqr}. Figure \ref{fig:scheme}a) shows a schematic of the experiment. 

Pairs of single photons were created via type-II spontaneous parametric down-conversion in a, nonlinear, potassium titanyl phosphate (PPKTP) crystal. This crystal was pumped by a $5$~mW diode laser centered at $410$~nm and emitted orthogonally polarized photon pairs with a wavelength of $820$~nm and a FWHM bandwidth of $0.6$~nm. The pairs were separated at a polarizing beamsplitter; one photon from each pair served as a trigger, while the second photon was launched into the QW. At an average heralded photon rate of ${\sim}20000$~s$^{-1}$, the mean longitudinal distance between two photons was about $250000$ times longer than the setup length of ${\sim}60$~cm. The probability of randomly creating more than one photon pair simultaneously was ${\sim}9\cdot10^{-5}$, i.e., only one photon was in the setup at any given time.

Quantum coin states were encoded in the polarization $\ket{H}$ and $\ket{V}$ of the input photon. Throughout our experiment, the initial coin state was set to left-circular polarization, $\ket{L}{=}(\ket{H}{+}i\ket{V})/\sqrt{2}$, using a quarter- and a half-wave plate, leading to symmetric probability distributions. For the results presented here, the remaining coin operators $C$ were Hadamards, realised with half-wave plates set to $22.5^\circ$. We can, however, prepare arbitrary pure input states as well as arbitrary coin operators for each step with suitable wave plate settings.

The lattice sites of the QW were represented by longitudinal spatial modes. The shift operator $S$ acting on these modes was implemented by a $27$~mm long, birefringent calcite beam-displacer. The displacers had a clear aperture of $20{\times}10~\mathrm{mm^2}$ and were  mounted on manual, tip-tilt rotation stages with a resolution of $217\;\umu \mathrm{rad}/5^{\circ}$ turn. The optical axis of each calcite prism was cut so that vertically polarized light was directly transmitted and horizontal light underwent a $2.7$~mm lateral displacement into a neighboring mode. Lattice sites were, typical for discrete walks on a line, labeled so that there were odd sites at odd time steps and even sites at even times.

The first two steps of the QW are shown in detail in Fig.~\ref{fig:scheme}b). The spatial modes after step $1$ were recombined interferometrically in step $2$. A series of steps then formed an interferometric network, Fig.~\ref{fig:scheme}a). We aligned this network iteratively, to a single interferometer per step. For example, the second beam displacer was aligned to maximize the interference visibility of the interferometer in Fig.~\ref{fig:scheme}b): the state $\ket{D}$ was input in mode 0 and the beam displacer rotated to maximize the overlap of the output mode 0 with $\ket{D}$. The third displacer was then aligned to the second and so on. We reached interference visibilities of typically $\sim99.8\%$ per step.

The photons emerging in the $N{+}1$ spatial modes at the output of an $N$-step QW were coupled into an optical fiber and subsequently detected by a single-photon avalanche photodiode, in coincidence with the trigger photon. We measured the probability distributions sequentially, translating the fiber coupler between the individual modes using a manual translation stage.

\begin{figure*}[ht!]
\begin{centering}
\includegraphics[width=\textwidth]{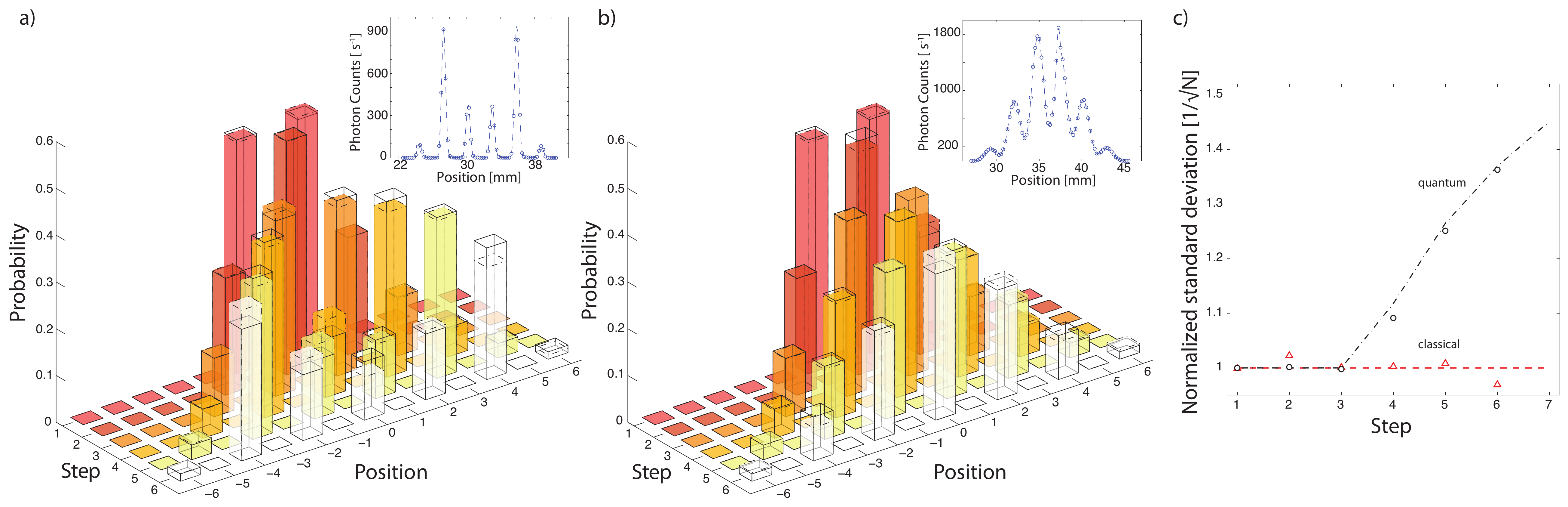}
\caption{\label{fig:walk}Probability distributions for successive steps of the (a) quantum  and (b) fully decohered (classical) walks up to the sixth step. Dashed lines show experimental data and solid lines show theoretical predictions. Probabilities are obtained by normalizing photon counts at each position to the total number of counts for the respective step. The insets show horizontal scans across the walk lattice for the five-step quantum walk (coupled into single-mode fiber) and decohered random walk (multi-mode fiber), respectively. (c) Normalized standard deviation of the probability distribution for quantum (black circles) and classical (red triangles) walks for $1$ to $6$ steps. Lines show the theoretical values; error bars are smaller than symbol size.}
\end{centering}
\end{figure*}

The measured probability distributions for detecting the photon at a given site, for 1 to 6 steps, are shown in Fig.~\ref{fig:walk}a). The experimental data are in excellent agreement with theory, with an average $L_1$-norm distance, $d=\frac12\sum_i\left|p_{i}^{exp}-p_{i}^{th}\right|$, of $0.031$ for the coherent and $0.017$ for the decohered walks. The quality of our data degrades somewhat for a higher number of steps, largely due to nonplanar optical surfaces, which caused small relative phase shifts between the multiple interferometers. The decohered walk is insensitive to phase errors and therefore better agrees with theory. 

Our scheme has several advantages: first, the interferometric network is inherently stable. The transversal mode-match is fulfilled because two beams emerging from one displacer will always be parallel, independent of small deviations in the optical alignment. The stability and alignment procedure of the QW grid are facilitated by the fact that the $N$ interferometers between steps $N$ and $(N{+}1)$ are formed between only $2$ optical components. Our setup---even though it is interferometric---does therefore not require active phase locking. Secondly, our system scales well, with the number of optical components increasing as $2N$ (as opposed to $(N^2{+}N)/2$ in \cite{jeong2004sqr,Do:05}), and exhibits low optical loss of $\sim 1\%$ per step. The remaining obstacle to scalability are nonideal optical components, a problem that can be alleviated with careful manufacturing or the use of shorter displacers. 

A unique feature of this setup is that tunable decoherence can be introduced by intentional misalignment of QW steps, Fig.~\ref{fig:scheme}c). Setting a non-zero relative angle between neighboring beam displacers leads to both a temporal delay and a transversal mode mismatch between interfering wave packets. Because the coincidence time window was much longer than the temporal shift we essentially integrated over the timing information, which corresponds to dephasing, cf.~Eq.~\ref{eq:deco}. Similarly, we traced over the spatial mode information by coupling the photons into a multi-mode fiber---as opposed to the single-mode fiber used for the coherent walks shown in Fig.~\ref{fig:walk}a). In practice, this reduced the interference contrast in all interferometers in the respective step. The QW was fully decohered [$q{=}1$ in Eq.~\ref{eq:deco}] when the interference visibility in each individual step reduced to $0$, which occured at a relative angle of $10.5^\circ$ in our experiment. Figure~\ref{fig:walk}b) shows the experimental results given by our system at full decoherence for steps $1$ to $6$. The probabilities---as expected for a classical random walk---follow a binomial distribution around the origin.

A distinguishing feature of an ideal QW is the speed at which the walker traverses the line. In particular, the standard deviation of the QW is proportional to the number of steps and not, as for the classical walk, its square root~\cite{kempe_quantum_2003}. This has been exploited to design quantum-walk--based search algorithms that exhibit a Grover-like quadratic speedup~\cite{shenvi_quantum_2003}. The measured standard deviations for both our quantum and fully decohered walks are shown in Fig.~\ref{fig:walk}c).  The results show very good agreement with theory: the fully decohered walk spreads diffusively, while the quantum walk spreads ballistically. 

\begin{figure}[b]
\includegraphics[width=.95\columnwidth]{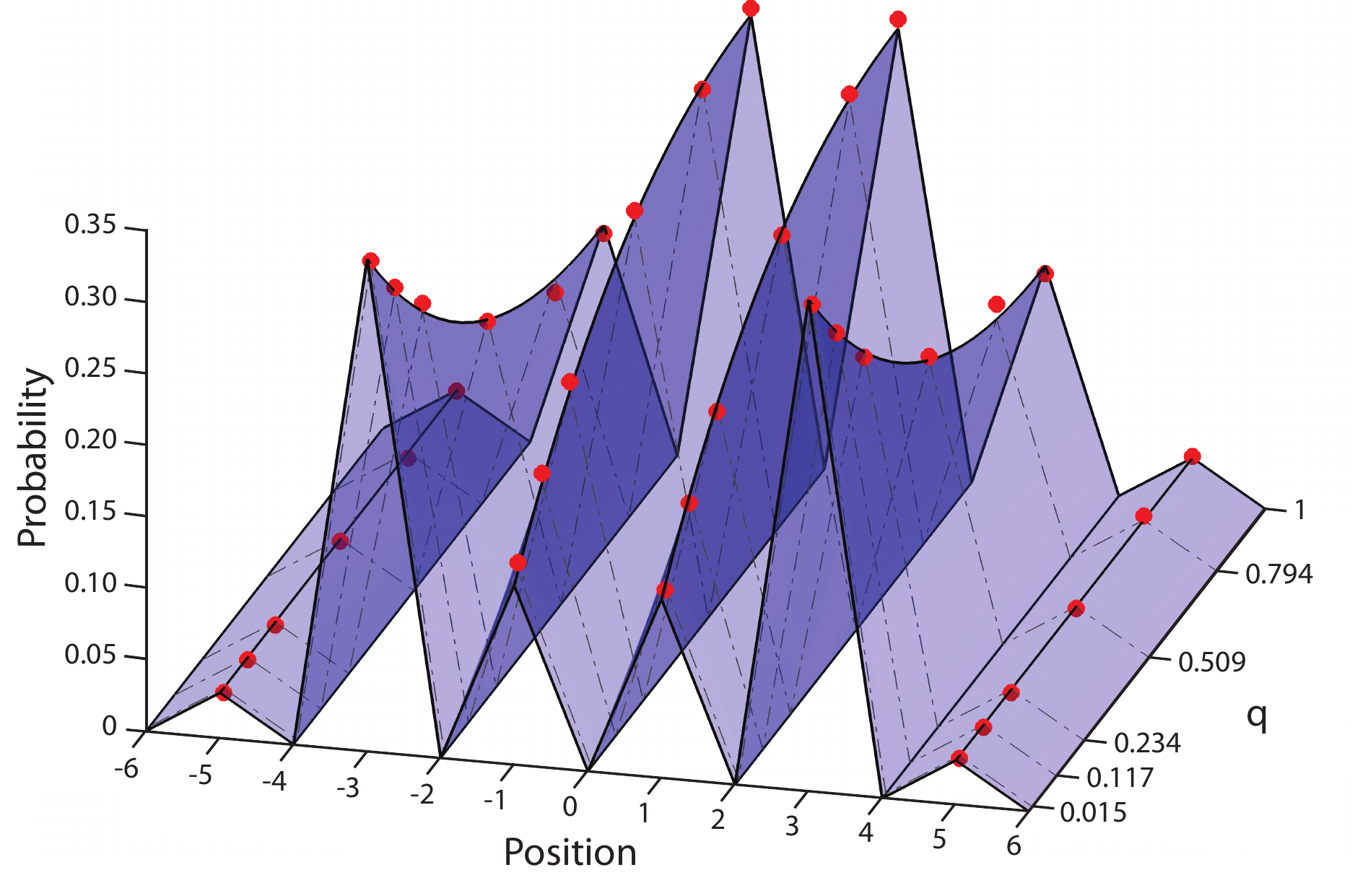}
\caption{\label{fig:transition}Transition of a 5-step quantum walk to a classical random walk. The decoherence parameter $q$ for a given data set was obtained from a least-squares fit to theory, Eq.~\ref{eq:deco}. All error bars are smaller than symbol size; the shaded area represents the theoretical probability distributions.}
\end{figure}

Tunable decoherence enabled us to investigate the quantum-to-classical transition for a $5$-step walk. By applying Eq.~\ref{eq:deco} to a two-step walk, see Fig.~\ref{fig:scheme}b), we calculated the interference visibility in output mode 0 after the second beam-displacer as a function of the decoherence $q$. We then adjusted the relative angle, Fig.~\ref{fig:scheme}c), between beam displacers to a target visibility. Figure~\ref{fig:transition} shows the resulting probability distributions, compared to theory, Eq. \ref{eq:deco}. Note the interesting nonlinear dependences of the probability distributions on $q$.

\begin{figure}[!t]
\includegraphics[width=0.9\columnwidth]{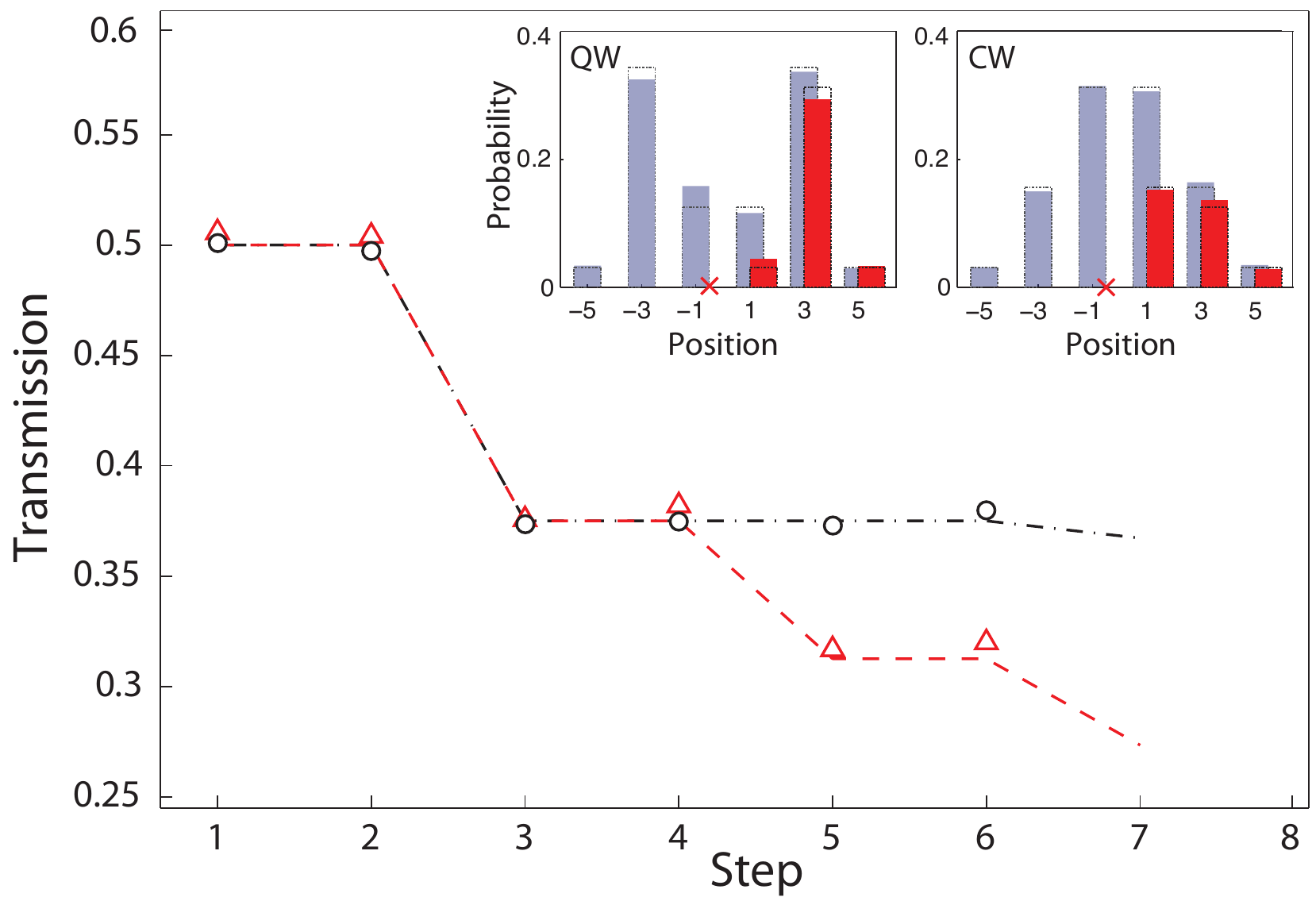}
\caption{\label{fig:absorption}Transmission probability of quantum (black circles) and classical (red triangles) walkers with an absorber located at position $-1$ and initial coin state $\ket{L}$. The transmission 
was obtained as the ratio of the number of transmitted photons measured with absorbers in place to the number measured without them. Error bars are smaller than symbol size. The insets show the fifth step walk with an absorber at position $-1$ (red columns) compared to the original walk (blue columns) for the quantum (QW) and classical case (CW).}
\vspace{-0.5 cm}
\end{figure}

Finally, we demonstrate another qualitative difference between classical and quantum walks, by incorporating absorbing boundaries. While a classical walker is eventually absorbed in the presence of an absorbing boundary, a quantum walker escapes with probability $1-2/\pi$~\cite{ambainis_one-dimensional_2001,bach_one-dimensional_2004}. A difference between the two exit probabilities first occurs after 5 steps, making it experimentally accessible with our current setup and providing a novel way of characterizing the degree of coherence in the walk. Absorbing boundaries were implemented using beam blocks in every spatial mode $-1$. Figure \ref{fig:absorption} shows the measured single-photon transmission probabilities in the quantum and fully decohered (classical) cases.

The most compelling features of our scheme are the ability to add tunable decoherence to a QW and the fact that every individual lattice site is fully accessible at any given time step. Future work could be to implement random or position-dependent coin operators to study walks on random environments \cite{yin2008qwr}, inhomogeneous walks \cite{linden2009iqw} and topological insulators \cite{kitagawa2010etp}. We could also add another walker on a separate line, using the existing setup with a vertical offset between input beams, or on the same line, with two or more photons launched in the same (or neighboring) spatial modes. This would allow exploration of entangled QW's~\cite{omar_quantum_2006}, as high-quality polarization-entangled photons can be routinely produced at high rates \cite{fedrizzi2007wtf}.  Finally, the setup can be used to prepare photon-number and path-entangled states across a large number of modes~\cite{papp2009cme}.

\begin{acknowledgments}
We thank the ARC Discovery and Federation Fellow programs and a IARPA-funded ARO contract. M.~A.~B acknowledges support by a UQRS/UQIRTA scholarship. I.~K. and A.~A.-G. thank the Dreyfus and Sloan Foundations, the ARO W911-NF-07-0304 and DARPA's Young Faculty Award program N66001-09-1-2101-DOD35CAP.
\end{acknowledgments}

\end{document}